\documentclass[pra,aps,twocolumn,showpacs,superscriptaddress]{revtex4-2}

\usepackage{amsmath,amssymb,graphicx}
\usepackage{float}
\usepackage{comment}
\usepackage{bm}
\usepackage{bbold}

\usepackage{graphicx}
\usepackage{dcolumn}
\usepackage{bm}
\usepackage{hyperref}
\usepackage{xcolor}
\usepackage{braket}
\usepackage{textcomp}
\usepackage{comment}
\usepackage{inputenc}
\usepackage{orcidlink}
\usepackage{algorithm}
\usepackage{algpseudocode}

\usepackage{quantikz}

\begin{document}

\title{Deterministic generation of grid states with programmable nonlinear bosonic circuits}

\author{Y. Le Fur~\orcidlink{0009-0000-8917-6923}}
\thanks{These authors contributed equally}
\email{yanis.lefur@csic.es}
\affiliation{Institute of Fundamental Physics IFF-CSIC, Calle Serrano 113b, 28006 Madrid, Spain}
\affiliation{Departamento de Física Téorica de la Materia Condensada and Condensed Matter Physics Center (IFIMAC),
Universidad Autónoma de Madrid, 28049 Madrid, Spain}

\author{J. Lalueza-Puértolas~\orcidlink{0009-0004-8988-6596}}
\thanks{These authors contributed equally}
\email{Javier.Lalueza@uab.cat}
\affiliation{Física Teòrica: Informació i Fenòmens Quàntics, Department de Física, Universitat Autònoma de Barcelona, 08193 Bellaterra (Barcelona), Spain}

\author{C. Sánchez Muñoz~\orcidlink{0000-0001-8775-0135}} 
\affiliation{Institute of Fundamental Physics IFF-CSIC, Calle Serrano 113b, 28006 Madrid, Spain}
\author{A. Mu$\tilde{\mathrm{n}}$oz de las Heras\,\orcidlink{0000-0003-3836-6107}}
\email{Alberto.MunozHeras@uclm.es}
\affiliation{Departamento de Física, Facultad de Ciencias Ambientales y Bioquímica, Universidad de Castilla-La Mancha, 45004 Toledo, Spain}
\author{A. González-Tudela\,\orcidlink{0000-0003-2307-6967}}
 \affiliation{Institute of Fundamental Physics IFF-CSIC, Calle Serrano 113b, 28006 Madrid, Spain}

\date{\today}

\begin{abstract}
Bosonic quantum error correction enables hardware-efficient protection of quantum information by encoding logical qubits in harmonic oscillators. Bosonic grid states, such as Gottesman–Kitaev–Preskill (GKP) states, are particularly promising due to their potential to correct small displacements and boson loss. However, their generation remains challenging, typically relying on probabilistic protocols or auxiliary qubit systems. Here, we propose deterministic protocols for generating bosonic grid states using programmable nonlinear bosonic circuits composed solely of squeezing, displacement, and Kerr operations.
We show that aiming to enforce GKP symmetries in the output of these circuits yields states with competitive performance with respect to current realizations, but whose quality saturates with increasing circuit depth due to imperfect symmetry restoration. Instead, we find that these bosonic circuits naturally give rise to a distinct class of states, that we label as phased-comb states, which are unitarily related to standard grid states but feature an intrinsic phase structure. We demonstrate that these states define a scalable bosonic quantum error-correcting code with near-optimal performance under boson loss comparable to that of approximate GKP states. We further analyze their logical operations and show how to implement a universal gate set for them. Our results establish programmable nonlinear bosonic circuits as a viable route towards the generation of scalable bosonic quantum error-correcting states beyond standard GKP encodings.
\end{abstract}

\maketitle

\section{Introduction}

Bosonic quantum error correction (QEC) encodes logical quantum information in the infinite-dimensional Hilbert space of harmonic oscillators, offering an alternative to discrete-variable architectures that require large qubit overheads~\cite{Shor1995,Steane1996,gottesman1997stabilizercodesquantumerror,Terhal2015}. Since the seminal proposal of Gottesman, Kitaev, and Preskill (GKP)~\cite{Gottesman2001}, the field has evolved with the introduction of a variety of bosonic codes, often tailored to specific hardware platforms and error models, highlighting the interplay between physical implementations and code design~\cite{Grimsmo2021,BRADY2024,Leghtas2013,Michael2016,xu2024lettingtigercagebosonic}. Among these, bosonic grid states, including GKP codes~\cite{Gottesman2001}, are particularly attractive due to their potential to correct small displacement errors and photon loss, that is the dominant noise mechanism in photonic platforms~\cite{Albert2018,Leviant2022quantumcapacity,Knill2001,Reagor2013,Slussarenko2019,Joshi2021}. Beyond their error-correcting capabilities, these states also find applications in quantum metrology, communication, and measurement-based quantum computing~\cite{Zhuang2020,labarca2025quantumsensingdisplacementsstabilized,Rozpedek2021,Menicucci2014,Larsen2021}. However, their non-Gaussian nature~\cite{Walschaers2021} makes their preparation a central challenge in bosonic quantum information.

Existing approaches to generate these states in the optical regime typically rely on measurements and post-selection~\cite{Thompson1992,Lodahl2015,SanchezMunoz2018,Delteil2019,Munoz-Matutano2019,Pirandola2004,Vasconcelos2010,Motes2017,Eaton2019,Su2019,Tzitrin2020,Larsen2025,Weigand18,takase2023gottesman}. Despite significant advances~\cite{Larsen2025}, most of these protocols suffer from a decreasing success probability with increasing photon number. Other approaches use hybrid architectures involving auxiliary systems, such as qubits, either to engineer dissipation and stabilize the target code autonomously~\cite{Royer2020,Campagne-Ibarcq2020,Rymarz2021,Sivak2023,Lachance-Quirion2024,Brock2025,Sellem2025}, or to deterministically generate them~\cite{Terhal2016,Shi2019,Kudra2022,Eickbusch2022,Kolesnikow2024,jacob2021measurement}. However, the additional complexity introduced by these auxiliary systems still limits the state-of-the-art GKP generation fidelities to $\lesssim 0.98$ with photon numbers $\lesssim 10$~\cite{Campagne-Ibarcq2020,Kudra2022,Eickbusch2022}. These results leave open the question of whether bosonic grid states can be generated through a protocol that is at once deterministic, scalable, and based solely on bosonic operations.

Motivated by recent experimental progress in programmable bosonic circuits~\cite{Cui2022,nielsen2024programmable,yang2025programmable,slim2025programmable,Zhang2017,Frattini2017,Kounalakis2018,Sivak2019,Ye2021,Brock2021nonlinear,He2023}, in this work we introduce a deterministic protocol for generating bosonic grid states using only squeezing, displacement, and Kerr nonlinear operations. Within this programmable framework, we identify two distinct approaches. In the first one, we aim to enforce the GKP translational symmetries in phase space into the generated states, leading to approximate GKP states with competitive photon numbers and fidelities with respect to state-of-the-art approaches~\cite{Campagne-Ibarcq2020,Kudra2022,Eickbusch2022}. However, we find that the quality of the symmetry enforcement saturates with increasing circuit depth, which ultimately limits scalability. In contrast, by relaxing the symmetry constraints and targeting only the grid structure, the protocol naturally generates a new class of states, which we label as \emph{phased-comb states}, that are unitarily related to standard grid states, but exhibiting an intrinsic phase structure. Using the near-optimal channel fidelity~\cite{Zheng2024,Zheng2025} as a benchmark, we show that these phased-comb states achieve scalable performance under boson loss comparable to that of approximate GKP states, and characterize their generation robustness under imperfect gate control and errors. Finally, we analyze how the phase structure of phased-comb states impacts logical operations and show how to construct a universal gate set, with most gates implemented using the same bosonic resources.

The manuscript is organized as follows. In Sections~\ref{sec:resources} and~\ref{sec:protocols}, we introduce the resources and protocols, respectively, used for generating both symmetry-enforced and phased-comb states. In Section~\ref{sec:code}, we analyze their performance as QEC codes, and then discuss the implementation of a universal gate set for the phased-comb states in Section~\ref{sec:gates}. Finally, Section~\ref{sec:conclusions} summarizes our results and outlines future directions.

\section{Programmable bosonic resources~\label{sec:resources}}
 
In this work, we consider the generation of bosonic grid states in a single mode described by annihilation (creation) operators $\hat{a}$ ($\hat{a}^\dagger$), with quadrature operators $\hat{x}=(\hat{a}+\hat{a}^\dagger)/\sqrt{2}$ and $\hat{p}=i(\hat{a}^\dagger-\hat{a})/\sqrt{2}$, satisfying $[\hat{x},\hat{p}]=i$. This mode can describe from localized or propagating photonic modes in the optical or microwave regimes, to other bosonic excitations such as phonons in motional degrees of freedom. However, for the sake of generality, in this work we adopt a platform-agnostic description. Our goal is then to generate the states assuming only the access to the following bosonic unitary operations:
\begin{align}
&\hat{U}_D(\alpha)=e^{\alpha \hat{a}^\dagger-\alpha^* \hat{a}}\,,\label{eq:disp}\\
&\hat{U}_S(r)=e^{\frac{r}{2}\left(\hat{a}^2-\hat{a}^{\dagger 2}\right)}\,,\label{eq:sq}\\
&\hat{U}_K(\chi t)=e^{-i\chi t \left(\hat{a}^\dagger \hat{a}\right)^2}\,,\label{eq:Kerr}
\end{align}
corresponding to displacement $\alpha$, squeezing $r$, and a Kerr-type nonlinear evolution during a time $t$ with a Kerr strength $\chi$, respectively. While displacement and squeezing are Gaussian operations readily available across bosonic platforms, the implementation of Kerr nonlinearities typically relies on coupling the bosonic mode to an underlying nonlinear element, such as a qubit, quantum emitter, or nonlinear medium. In contrast to approaches where such auxiliary systems are actively used for control or state generation~\cite{Royer2020,Campagne-Ibarcq2020,Rymarz2021,Sivak2023,Lachance-Quirion2024,Brock2025,Sellem2025,Terhal2016,Shi2019,Kudra2022,Eickbusch2022,Kolesnikow2024}, we consider a strategy in which only the effective Kerr evolution of the bosonic mode is exploited, without requiring active hybrid control during the protocol.

\section{Deterministic Grid State Generation Protocols~\label{sec:protocols}}

The bosonic state generation protocols considered in this work start from the bosonic squeezed vacuum state, $\ket{\Psi_0}=\hat{U}_S(r)\ket{0}$, and apply to it a nonlinear bosonic circuit concatenating the operations defined in Eqs.~\eqref{eq:disp}-\eqref{eq:Kerr}, i.e.,
\begin{equation}
\ket{\Psi_\mathrm{out}}=\prod_i \hat{U}_K(\chi_i t_i)\hat{U}_D(\alpha_i)\ket{\Psi_0}\,.
\end{equation}

Our objective is to identify circuit parameters $(\alpha_1,\chi_1 t_1,\dots)$ such that the generated state approximates a bosonic grid state with desirable QEC properties.

Ideally, one would like to generate the ideal GKP code states~\cite{Gottesman2001}, which exhibit a periodic grid structure in phase space and are stabilized by translational symmetries. In practice, however, these ideal states have infinite energy and must be approximated by finite superpositions of displaced squeezed states, for instance, through Gaussian-envelope~\cite{Gottesman2001} or hard cut-off truncations with a finite number of ``legs", labeled as \emph{comb states}~\cite{Brenner2025}. Precise definitions of all these states are provided in Appendix~\ref{app:definition}.

In what follows, we explain the two distinct approaches for generating grid states within our circuit model: one where we attempt to enforce the translational symmetries characteristic of GKP states, that we discuss in Section~\ref{subsec:symmetry}, and another one where we relax these symmetry constraints and instead target the grid structure itself, which we explain in Section~\ref{subsec:phased-comb}.

\subsection{Symmetry-enforced states~\label{subsec:symmetry}}

\begin{figure*}[hbt!]
    \centering
   \includegraphics[width=\linewidth]{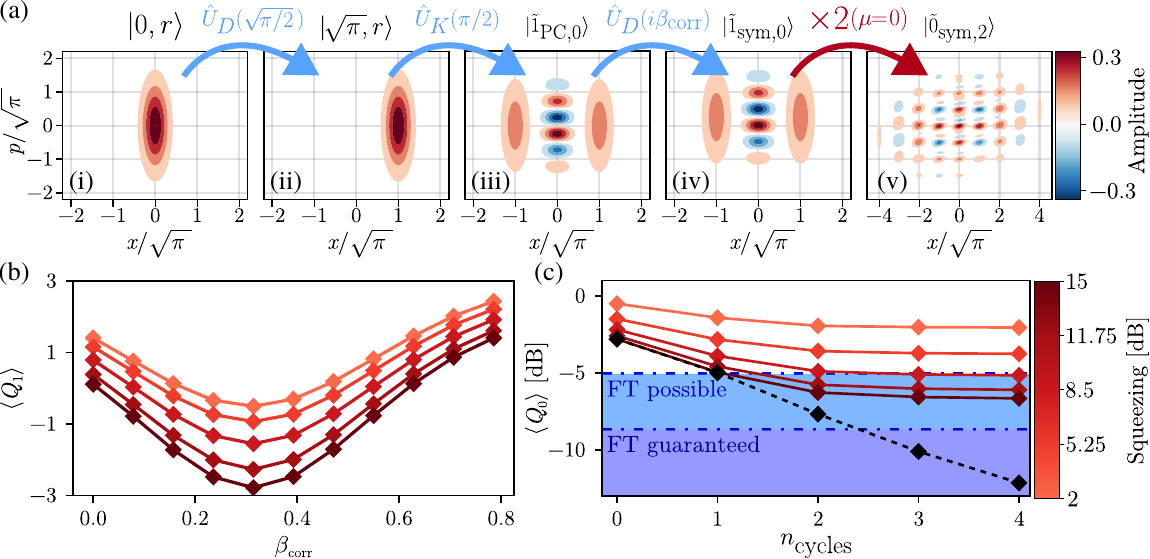}
    \caption{\textbf{Symmetry-enforced state preparation for the $0$-logical state.} (a) Wigner representations of the initial squeezed vacuum state with $r = 6$ dB (i), displacement (ii), Kerr evolution (iii), and a final correction displacement $\hat{U}_D(i\beta_\text{corr})$ (iv). Panel (v) represents the Wigner after applying three cycles of the protocol described in Algorithm~\ref{alg:symmetrygkp} with initial squeezing $r=6$ dB. (b) Expectation value $\braket{\hat{Q}_1}$ [dB] versus correction amplitude $\beta_\text{corr}$ for a single-cycle generated state and for different values of initial squeezing $r$ in different red shades. (c) $\braket{\hat{Q}_0}$ [dB] saturation for an increasing number of cycles of the symmetry-enforced states across different squeezing regimes (in red shades). The performance is compared against a comb state~\cite{Brenner2025} with $r=15$ dB (black) and the same number of legs $2^{n_\mathrm{cycles}+\mu}+(1-\mu)$. Due to the convention chosen for the initial step, for $n_\mathrm{cycles}=0$ we plot $\braket{\hat{Q}_1}$ Values are calculated as $\braket{\hat{Q}_\mu}$ [dB] $= 10\log_{10}(\braket{\hat{Q}_\mu})$.}
    \label{fig:state_preparation}
\end{figure*}

To construct approximate GKP states within our circuit model, let us first analyze the effect of a single circuit cycle on the phase-space symmetries of the bosonic mode. As shown in Figs.~\ref{fig:state_preparation}(a.i)-(a.iii), a cycle consists of applying displacement and Kerr operations to an initial squeezed vacuum state. The corresponding Wigner distributions of the states show that the combination of squeezing and displacement generates a displaced squeezed state, $\ket{\alpha,r}$, while the Kerr evolution with $\chi t=\pi/2$ produces a superposition of two displaced squeezed components located at $\pm \alpha$. Crucially, the Kerr operation introduces a relative phase between the two lobes, yielding
\begin{equation}
\hat{U}_{K}(\pi/2)\ket{\alpha,r}=\frac{1}{\sqrt{2}}\left(e^{-i\frac{\pi}{4}}\ket{\alpha,r}+e^{i\frac{\pi}{4}}\ket{-\alpha,r}\right).
\end{equation}
This relative $e^{i\pi/2}$-phase breaks the reflection symmetry of the interference fringes in phase space, as shown in Fig.~\ref{fig:state_preparation}(a.iii), deviating from the ideal GKP structure.

To partially restore this symmetry, we introduce a correction step consisting of an extra displacement $\hat{U}_D(i\beta)$ after the initial operations. The parameter $\beta$ is chosen to minimize one of the GKP squeezing operators defined in Ref.~\cite{marek2024} for the two logical GKP states, reading 
\begin{align}
\hat{Q}_{1 (0)}=\frac{1}{2}\left(4+(-)\sqrt{\hat{S}_p}+(-)\sqrt{\hat{S}_p^\dagger}-\hat{S}_x-\hat{S}_x^\dagger\right)\,,\label{eq:Q0sq}
\end{align}
where $\hat{S}_x=e^{-2i\sqrt{\pi}\hat{p}}$ and $\hat{S}_p=e^{-2
i\sqrt{\pi}\hat{x}}$ are the stabilizers of the GKP code. Since the ideal GKP state $\ket{0_\text{GKP}}$ ($\ket{1_\text{GKP}}$) is a ground state of $\hat{Q}_0$ ($\hat{Q}_1$) with zero eigenvalue, minimizing $\langle \hat{Q}_0 \rangle$ ($\langle \hat{Q}_1 \rangle$) with $\beta$ should lead to a symmetry restoration of these states. For example, in Fig.~\ref{fig:state_preparation}(b), we plot the $\braket{\hat{Q}_1}$ of the state $\hat{U}_D(i\beta)\hat{U}_K( \pi/2)\ket{\alpha,r}$ as a function of $\beta$, showing how it features a minimum for the $\alpha,r$ parameters chosen around $\beta_\mathrm{corr}\approx 0.31$. Applying this correction to the states after the three initial operations described in Figs.~\ref{fig:state_preparation}(a.i)-(a.iii) improves the symmetry of the interference fringes, as illustrated in the Wigner distribution of Fig.~\ref{fig:state_preparation}(a.iv), where the corrected state closely resembles the reflection symmetry of the ideal GKP structure.

Based on this mechanism, we propose an iterative protocol to generate large-scale symmetry-enforced logical states by concatenating layers of these operations, see Algorithm~\ref{alg:symmetrygkp}. The protocol begins with a common initialization and correction step ($j=0$), as described above, after which the logical state is selected through the structure of the subsequent iterations. The prepared state after initialization is considered as the $1$-logical state ($\mu = 1$) with two legs. After that, the parameter $\mu \in \{0,1\}$ determines whether the first leg-doubling step is included, which in turn fixes the offset of the resulting grid in phase space. Each iteration consists of two steps. First, a displacement followed by a Kerr operation generates additional components, effectively doubling the number of legs of the state. Second, an additional displacement is applied to partially restore the GKP symmetry disrupted by the Kerr evolution. While successive iterations increase the number of legs, they preserve the underlying grid offset determined by $\mu$. We denote the resulting states after $j$ cycles as $\ket{\tilde{\mu}_{\mathrm{sym},j}}$. 
Note that these states feature a hard cut-off truncation and thus can be directly compared with comb states~\cite{Brenner2025}.

\begin{algorithm}[H]
\caption{Protocol to generate grid states}\label{alg:symmetrygkp}
\label{alg:protocol}
\begin{algorithmic}[1]
\Require Initial squeezed vacuum state $\ket{\Psi_0}=\hat{U}_S(r)\ket{0}$, total cycles $n_{\rm cycles}$, logical state $\mu \in \{0, 1\}$
\State $\ket{\Psi} \gets \hat{U}_{\rm K}(\pi/2) \hat{U}_D(\sqrt{\pi/2})\ket{\Psi_0}$ \Comment{Initialization}
\State $\ket{\Psi} \gets \hat{U}_{D}(i\beta_\mathrm{corr})\ket{\Psi}$ \Comment{Correction optimizing $\braket{Q_1}$}
\For{$j = 1$ to $n_{\rm cycles}$}
    \State $\ket{\Psi} \gets \hat{U}_{K}(\pi/2)\hat{U}_D(2^{j-1+\mu}\sqrt{\pi/2})\ket{\Psi}$  \Comment{Leg doubling}
    \State $\ket{\Psi} \gets \hat{U}_{D}(i\beta_\mathrm{corr})\ket{\Psi}$  \Comment{Correction optimizing $\braket{Q_\mu}$}
\EndFor
\State \Return $\ket{\Psi}$
\end{algorithmic}
\end{algorithm}

In Fig.~\ref{fig:state_preparation}(c), we plot the evolution of the GKP squeezing, $\braket{\hat{Q}_0}$, of the $0$-logical generated states $\ket{\tilde{0}_{\mathrm{sym},n_\mathrm{cycles}}}$\footnote{Let us note that due to our convention chosen, we plot $\braket{Q_1}$ when $n_\mathrm{cycles}=0$} following this protocol as a function of the number of cycles ($n_\mathrm{cycles}$) and for different values of the initial squeezing (shown in different shades of red). For comparison, we also include the value of $\langle \hat{Q}_0 \rangle$ for comb states defined in Ref.~\cite{Brenner2025}(black markers) at the largest squeezing considered ($r=15$ dB), constructed with the same number of legs, i.e., $2^{n_\mathrm{cycles}+\mu}+(1-\mu)$. 

We observe that, irrespective of the initial squeezing, $\langle \hat{Q}_0 \rangle$ for the symmetry-enforced states decreases initially but rapidly saturates after a few cycles. This reflects the inability of the correction step to fully compensate the phase mismatch introduced by the Kerr evolution. In contrast, the ideal comb states approach $\langle \hat{Q}_0 \rangle \rightarrow 0$ monotonically with increasing number of cycles (or equivalently, number of legs), highlighting that the limitation arises from imperfect symmetry enforcement rather than from the grid structure itself. 

Increasing the squeezing further improves the achievable values of $\langle \hat{Q}_0 \rangle$, approaching the fault-tolerant threshold defined in Ref.~\cite{marek2024,Bourassa2021blueprintscalable}. Moreover, already for a small number of cycles $n_\mathrm{cycles}=3$ and for state-of-the-art squeezing values $r=7.8$ dB~\cite{Eickbusch2022,Vaartjes2024}, the protocol generates a state with approximate GKP symmetries [see Fig.~\ref{fig:state_preparation}(a.v)] with a fidelity $\mathcal{F}_\mathrm{comb}\approx 95.1\%$ with respect to the comb states defined in Ref.~\cite{Brenner2025}, competitive with current experimental realizations~\cite{Campagne-Ibarcq2020,Kudra2022,Eickbusch2022}. 

\subsection{Phased-comb states~\label{subsec:phased-comb}}

\begin{figure*}[htb!]
    \centering
   \includegraphics[width=\linewidth]{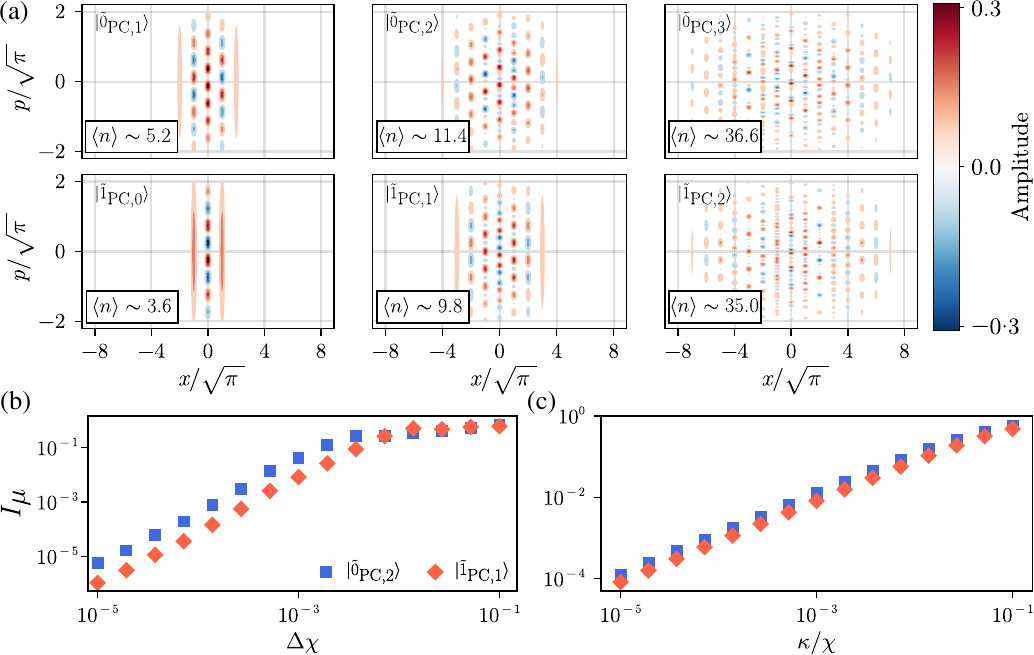}
    \caption{\textbf{Wigner distributions and logical state infidelities for phased-comb states.} (a) Wigner distributions of the phased-comb logical states $\ket{\tilde{0}_{\text{PC},j}}$ ($n_{\rm cycles}=1, 2$, and $3$ cycles, left to right) and $\ket{\tilde{1}_{\text{PC},j}}$ ($n_{\rm cycles}=0, 1$, and $2$ cycles, left to right) using a squeezing parameter of $r  = 10$ dB  and the protocol described in Algorithm~\ref{alg:protocol}, but without the correction step. (b,c) Logical state infidelity $I_\mu$ ($\mu\in\{0,1\}$) as a function of the Kerr uncertainty $\Delta{}\chi_\text{max}$ (b) and the boson loss rate normalized to the Kerr strength $\kappa/\chi$ (c). The infidelity is evaluated for a $3$-cycle state ($\mu = 0$, blue squares) and a 2-cycle state ($\mu = 1$, red rhombus) at a squeezing of $r = 7.8$ dB. Data points represent an average over 100 realizations; error bars indicating one standard deviation are smaller than the marker size.}
    \label{fig:state_phase}
\end{figure*}

We now consider the protocol described in Algorithm~\ref{alg:symmetrygkp}, but without the correction step. The circuit naturally generates two classes of states depending on the iteration structure controlled by $\mu$, which determines the offset of the resulting grid in phase space. In contrast to the symmetry-enforced protocol, these states preserve the grid structure for arbitrary circuit depth. As shown in Fig.~\ref{fig:state_phase}(a), the resulting states exhibit a comb-like structure in phase space whose number of legs doubles with the number of cycles, while maintaining coherence and orthogonality between them. This allows us to define two logical states corresponding to $\mu=0,1$ as phased comb states with $2^{n_\mathrm{cycles}}+1$ and $2^{n_\mathrm{cycles}+1}$ legs, respectively.

Crucially, the Kerr evolution introduces a nontrivial phase and amplitude structure across the legs, which distinguishes these states from standard comb or GKP states~\cite{Brenner2025}, and fundamentally modifies their symmetry properties. As shown in Appendix~\ref{app:generation_comb}, the phase and amplitude structure of the generated states is given by:
\begin{align}
\ket{\tilde{\mu}_{\mathrm{PC},j}} &= \sum_{s=-s_{\rm max}}^{s_{\rm max}}
e^{i\Phi[(s+\mu/2)2\sqrt{\pi}]}\delta_s(\mu,s_{\rm max})\nonumber\\
&\times\ket{(s+\mu/2)\sqrt{2\pi},r},
\label{eq:PC}
\end{align}
where $\Phi(x)$ and $\delta_s(\mu,s_{\rm max})$ are real functions determined by the circuit whose expression is explicitly given in Appendix~\ref{app:generation_comb}, and $s_\mathrm{max}=2^{j+1}$ is the 
number of legs in the $j$-th iteration. Note that the amplitude difference, $\delta_s(\mu,s_{\rm max})$, only occurs at the boundary legs of the $0$-logical comb state, and thus its importance vanishes with increasing number of cycles. Therefore, in the large number of cycles limit, the generated states can be shown to be unitarily related to the comb states introduced in Ref.~\cite{Brenner2025} via a position-dependent phase operator $\hat{U}=e^{i\Phi(\hat{x})}$, which is why we label them as phased-comb states. Since the grid structure of these states is preserved under ideal, noiseless operations, the protocol is intrinsically scalable in the absence of imperfections, unlike the symmetry-enforced states. We therefore assess the scalability of phased-comb states by considering the impact on their generation of imperfect gate control and non-unitary errors during the Kerr evolution, which is the most demanding experimental operation.

In particular, Fig.~\ref{fig:state_phase}(b) shows the generation infidelity under imperfect coherent control of the Kerr evolution, which we model as deviations $\chi'=\pi/2+\Delta\chi$, with $\Delta\chi$ uniformly sampled in $[-\Delta\chi_\text{max},\Delta\chi_\text{max}]$. The infidelity is defined as $I_\mu=1-|\langle \tilde{\mu}^{(\Delta \chi)}_{\rm PC}|\tilde{\mu}_{\rm PC}\rangle|^2$, where $\ket{\tilde{\mu}^{(\Delta \chi)}_{\rm PC}}$ is the state generated assuming an imperfect parameter choice against the ideal one $\ket{\tilde{\mu}_{\rm PC}}$. We find that maintaining errors below the $10\%$ level requires $\Delta\chi_\text{max}\lesssim 10^{-2}$, indicating a stringent requirement on Kerr control. In Fig.~\ref{fig:state_phase}(c), we analyze the effect of non-unitary errors during the Kerr evolution, e.g., due to the decay of the bosonic mode at a rate $\kappa$. Formally, we include boson losses by switching to the density matrix representation of quantum states and assuming a non-unitary time evolution given by a Lindblad master equation during a time $t\chi=\pi/2$. We plot the resulting infidelity as a function of $\kappa/\chi$, showing that achieving errors below $10\%$ requires $\kappa/\chi \lesssim 10^{-2}$, which is compatible with current experimental capabilities in microwave platforms~\cite{Grimm2020}.

\section{Quantum error correction with phased-comb states~\label{sec:code}}

\begin{figure}[t!]
    \centering
    \includegraphics[width=\linewidth]{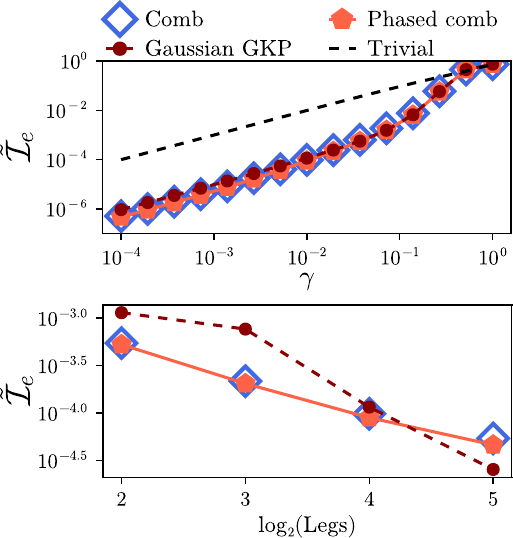}
    \caption{\textbf{Performance of QEC codes.} (a) Near-optimal channel infidelity $\mathcal{\tilde{I}}_e = 1-\tilde{\mathcal{F}}_e$ as a function of the photon loss probability $\gamma$. (b) Channel infidelity versus the number of state legs at a fixed loss probability $\gamma = 10^{-2}$. Performance is compared across four encodings: phased-comb (red rhombus), comb (blue rhombus), Gaussian-truncated GKP (red circles), and the trivial encoding $\{\ket{0},\ket{1}\}$ (dashed line). All  codes are evaluated with the same number of legs. Parameters: (a) $\mu=0$ logical states generated with $3$ cycles and $\mu=1$ logical states with $2$ cycles. (b) Truncation level of the cavity $N_R = 500$ except for the $5$ cycle case where $N_R = 1200$ to ensure convergence. (a,b) Squeezing is fixed at $r = 7.8$ dB. The GKP envelope is set by $\Delta = \exp(-5r)$.}
    \label{fig:channel_fidelity}
\end{figure}

We now show that phased-comb states define a scalable bosonic quantum error-correcting code with performance comparable to approximate GKP states with a Gaussian envelope truncation, while arising naturally from our circuit model. To quantify the error-correcting capabilities of these states, we evaluate the near-optimal channel fidelity $\tilde{\mathcal{F}}_{\rm e}$~\cite{Zheng2024,Zheng2025}, see Appendix~\ref{app:Channel_fidelity} for definitions and calculation details. This quantity provides an upper bound on the performance achievable with optimal recovery operations and for a given noise channel, making it a suitable benchmark for comparing different encodings. We compare the codes generated with phased-comb states $\{\ket{\tilde{0}_{\mathrm{ PC},j}},\ket{\tilde{1}_{\mathrm{ PC},j}}\}$ of Eq.~\eqref{eq:PC} with standard comb states~\cite{Brenner2025}, Gaussian-truncated GKP states~\cite{Gottesman2001}, and the trivial encoding $\{\ket{0},\ket{1}\}$. For a fair comparison, all codes are constructed using the same number of legs, corresponding to the same number of terms in their respective superpositions. The comb code is defined using Eq.~\eqref{eq:PC} removing the phase and amplitude structure, while Gaussian-truncated GKP states are obtained by applying a Gaussian envelope operator $\exp(-\Delta\hat{a}^\dagger\hat{a})$ to the comb states. Regarding the noise mechanism, we focus on bosonic loss, which is the dominant one in most bosonic platforms. We parametrize this noise by a parameter $\gamma=1-e^{-\kappa T}$, which is the boson loss probability within a time window $T$, assuming a boson decay rate $\kappa$.

Figure~\ref{fig:channel_fidelity}(a) shows the channel infidelity $\mathcal{\tilde{I}}_{\rm e}=1-\tilde{\mathcal{F}}_{\rm e}$ as a function of the boson loss probability $\gamma$. All non-trivial encodings provide a significant improvement over the trivial code. For small loss rates up to $\gamma \sim 10^{-2}$, comb and phased-comb states achieve slightly lower infidelities than the Gaussian-truncated GKP states. Moreover, in the moderate-to-large loss regime, all three bosonic codes exhibit comparable performance. Notably, phased-comb states match the performance of standard comb states across the entire range, despite lacking exact GKP symmetries. This demonstrates that enforcing exact translational symmetry is not necessary to achieve near-optimal protection against bosonic loss within this class of states. 

To assess the dependence on code size, Fig.~\ref{fig:channel_fidelity}(b) shows $\mathcal{\tilde{I}}_{\rm e}$ as a function of the number of legs for a fixed loss probability $\gamma=10^{-2}$. For small number of legs, comb and phased-comb codes exhibit slightly better performance. As the number of legs increases, Gaussian-truncated approximate GKP states achieve slightly lower infidelities, while comb and phased-comb states remain comparable to each other across the explored range. These results further support that the phase structure of phased-comb states does not significantly degrade their error-correcting performance under bosonic loss.

\section{Logical operations and measurement in phased-comb codes~\label{sec:gates}}

We now show that universal quantum computation can be performed within the phased-comb encoding, identifying the modifications required with respect to standard GKP codes due to the presence of an intrinsic phase structure. In Section~\ref{subsec:frame}, we show how most logical operations can be the same as standard GKP ones, except for the Hadamard gate. In addition, Section~\ref{subsec:distortion} discusses how the phase structure affects measurements in the $p$ quadrature.

\subsection{Logical operations in the phase frame~\label{subsec:frame}}

To understand how logical operations are implemented in the phased-comb code, it is convenient to relate it to the standard GKP encoding. For that, we take the ideal, infinite-energy limit, where the phased-comb codewords can be simply written as unitary transformations of GKP codewords, i.e., $\ket{\tilde{\psi}}=\hat{U}\ket{\psi}$, where $\hat{U}=e^{i\Phi(\hat{x})}$ and $\ket{\psi}$ denotes the GKP codewords. As a consequence, logical operators $\hat{O}$ acting on GKP states are mapped to operators $\hat{\tilde{O}}=\hat{U}\hat{O}\hat{U}^\dagger$ acting on phased-comb states. This transformation defines a \emph{phase frame}, that allows us to understand how the different logical operations can be applied with this encoding.

Within this framework, it becomes transparent that the effect of the phase operator on logical gates depends on their commutation properties with the position operator $\hat{x}$. Operators that commute with $\hat{x}$ remain unchanged under the transformation. This includes the logical $\hat{Z}$ operator, the stabilizer $\hat{S}_p$, and phase gates such as $\hat{S}$ and $\hat{T}$. These operations can therefore be implemented exactly as in the standard GKP code. In contrast, operators that depend on the momentum quadrature $\hat{p}$ are modified in the phase frame. For instance, applying the Pauli $\hat{X}$ operator and the stabilizer $\hat{S}_x$ of the GKP code to a phased-comb state leads to a shifted phase dependence, i.e., $\hat{U}'=\hat{U}\hat{O}\hat{U}^\dagger=e^{i\Phi(\hat{x}-\sqrt{\pi})}$ and $\hat{U}'=e^{i\Phi(\hat{x}-2\sqrt{\pi})}$, respectively, which can be absorbed into an updated phase frame without altering their physical implementation. This allows these operations to be performed using the same bosonic resources as in the GKP encoding.

A qualitatively different situation arises for the Hadamard gate $\hat{H}$, which for the GKP code implements a Fourier transform exchanging $\hat{x}\leftrightarrow\hat{p}$. Under this transformation, the phase operator becomes momentum-dependent, $\hat{U}'=\hat{H}\hat{U}\hat{H}^\dagger = e^{i\Phi(\hat{p})}$. This is problematic because momentum-dependent phases do not commute with the standard two-qubit entangling gate $\hat{\mathrm{CZ}}=e^{i\hat{x}_1\hat{x}_2}$. As a result, the phase structure spreads across different modes during entangling operations, effectively coupling the phase frames of different qubits and making error tracking intractable. To avoid this issue, we propose not to apply the GKP Hadamard gate directly. Instead, one can implement the Hadamard operation of the phased-comb code using an ancilla-assisted protocol based on gate teleportation (see Appendix~\ref{app:hadamard}). This approach performs the logical basis change while ensuring that the phase operator remains a function of $\hat{x}$, thereby preventing the uncontrolled propagation of phase correlations across the system.

\subsection{Measurement and quadrature distortion in the phase frame~\label{subsec:distortion}}

The presence of the phase operator $\hat{U}=e^{i\Phi(\hat{x})}$ also affects quadrature measurements, in a similar fashion to the logical gates discussed above. In particular, while measurements in the position quadrature $\hat{x}$ remain unaffected, measurements in the momentum quadrature $\hat{p}$ are distorted by the phase structure. This asymmetry originates from the fact that the phase operator is diagonal in the $\hat{x}$ basis. As a result, homodyne measurements of $\hat{x}$ directly probe the grid structure of the phased-comb states without modification. In contrast, measurements in the $\hat{p}$ quadrature involve a Fourier transform of the state, which converts the position-dependent phase into a nontrivial amplitude modulation.

This effect can be understood by analyzing the $\hat{p}$-quadrature eigenstates of the phased-comb code (see Appendix~\ref{app:momentum_eigenstates}), which take the form
\begin{equation}
    \begin{split}
        &\ket{\tilde{+}_{\rm PC}} = \sum_{k\in\mathbb{Z}} \int \frac{dp}{\sqrt{2\pi}} \tilde{f}(p - 2k\sqrt{\pi}) \ket{p},\\
        &\ket{\tilde{-}_{\rm PC}} = \sum_{k\in\mathbb{Z}} \int \frac{dp}{\sqrt{2\pi}} \tilde{f}(p - (2k+1)\sqrt{\pi}) \ket{p},\\
    \end{split}
\end{equation}
where the function $\tilde{f}(p)$ is the Fourier transform of the phase profile $e^{i\Phi(\hat{x})}$. In the absence of phase modulation, $\tilde{f}(p)$ reduces to a Dirac-delta function and one recovers the ideal GKP momentum eigenstates. However, for a nontrivial phase function $\Phi(\hat{x})$, $\tilde{f}(p)$ acquires a finite width, leading to a broadening and deformation of the legs in momentum space.

As a consequence, direct $\hat{p}$-quadrature measurements no longer provide sharp information about the logical state, limiting their usefulness for readout and error correction. To overcome this issue, momentum measurements can be effectively implemented by combining a phase-preserving Hadamard operation with a subsequent $\hat{x}$-quadrature measurement. In this way, both logical operations and measurements can be performed consistently within the phased-comb encoding while preserving the phase structure.

\section{Conclusions~\label{sec:conclusions}}

In this work, we introduce deterministic protocols for generating bosonic grid states using programmable nonlinear bosonic circuits based on squeezing, displacement, and Kerr operations. We identify two approaches: enforcing GKP symmetries yields states with competitive fidelities but limited scalability, while relaxing these constraints naturally produces phased-comb states that preserve the grid structure and define a scalable encoding with near-optimal protection against boson loss. Moreover, we demonstrate that universal quantum computation is possible within this encoding through a phase-frame description of logical operations. Most logical gates can be implemented with the same bosonic resources. For operations involving Fourier transforms, like Hadamard gates, we present an ancilla-assisted strategy that preserves the phase structure, demonstrating that these operations remain fully accessible within the encoding. A possible direction for future work is to investigate whether fully bosonic implementations of the complete logical gate set can be achieved, for example through optimal control techniques~\cite{Grochowski2025quantumcontrolof,kendell2024deterministic} or by engineering effective nonlinear interactions that preserve the phase structure. Additionally, extending this programmable framework to multimode settings or other bosonic code structures~\cite{Michael2016,Royer2022,Brock2025,xu2025letting} may reveal new families of scalable quantum error-correcting codes beyond conventional ones.

\section*{Code and Data Availability}
The data that support the findings of this study are available in \cite{lefur2026_code}. The numerical simulations were performed using \verb|QuantumToolbox.jl|~\cite{mercurio2025} and \verb|Optimization.jl|~\cite{vaibhav_kumar_dixit_2023_7738525}.

\begin{acknowledgements}
  The authors acknowledge support from the CSIC Research Platform on Quantum Technologies PTI-001. AGT acknowledges support from Spanish project Proyecto PID2024-162384NB-I00 financiado por MICIU/AEI/10.13039/501100011033 y por FEDER,UE, from the
QUANTERA project MOLAR with reference PCI2024153449 and funded MICIU/AEI/10.13039/501100011033
and by the European Union, the Programa Fundamentos FBBVA through the grant EIC24-1-17304. AMH acknowledges support from Fundación General CSIC's ComFuturo program, which has received funding from the European Union's Horizon 2020 research and innovation program under the Marie Skłodowska-Curie grant agreement No. 101034263. CSM and YLF acknowledge support by the project PID2023-149969NA-100 funded by the Spanish Agencia Estatal de Investigación MICIU/AEI/10.13039/501100011033, by a 2025
Leonardo Grant for Scientific Research and Cultural Creation from the BBVA Foundation, and by project NPhoQuss funded by Programa Fundamentos 2024 from the BBVA Foundation through the grant EIC24-1-17304.
\end{acknowledgements}

\appendix


\section{Definition of the different target states}
\label{app:definition}

Here, we briefly review the definitions of the bosonic grid states considered in this work. The ideal Gottesman–Kitaev–Preskill (GKP) code states~\cite{Gottesman2001} are defined as superpositions of position eigenstates:
\begin{equation}
\ket{\mu_{\rm GKP}} \propto \sum_{s \in \mathbb{Z}} \ket{(2s+\mu)\sqrt{\pi}}_{\hat{x}},
\label{eq:idealGKP}
\end{equation}
where $\mu=0,1$ labels the logical states. These states exhibit exact translational symmetry in phase space but have infinite energy and are therefore unphysical.

In practice, finite-energy approximations are required. A common construction is the Gaussian-truncated GKP state~\cite{Gottesman2001},
\begin{equation}
|\tilde{\mu}_{\rm GKP}\rangle \propto \sum_{s\in\mathbb{Z}} e^{-2\pi\Delta^2 \left(s+\frac{\mu}{2}\right)^2}
\hat{U}_D\left[(s+\mu/2)\sqrt{2\pi}\right]\hat{U}_S(r)\ket{0},
\label{eq:truncatedGKP}
\end{equation}
which preserves the grid symmetries while regularizing the energy. 

An alternative truncation is given by the comb states introduced in Ref.~\cite{Brenner2025},
\begin{equation}
|\tilde{\mu}_{\rm C}\rangle \propto \sum_{s=-s_{\rm max}}^{s_{\rm max}}
\hat{U}_D\left[(s+\mu/2)\sqrt{2\pi}\right]\hat{U}_S(r)\ket{0},
\label{eq}
\end{equation}
where $s_{\rm max}$ sets the number of legs of the superposition. These states retain the grid structure but lack the Gaussian envelope. They can be converted into Gaussian-truncated GKP states through purification protocols~\cite{Brenner2025}, and serve as the target states for the symmetry-enforced protocol discussed in Sec.~\ref{subsec:symmetry}.

\section{Generation of phased-comb states}
\label{app:generation_comb}

In this Appendix, we detail the origin and propagation of the phase structure generated by the Kerr nonlinearity in the phased-comb protocol. For that, we first consider the case of a single cycle in Section~\ref{subsec:single} and then study how it propagates within the protocol in Section~\ref{subsec:propagation}.

\subsection{Kerr-induced phase in a single cycle~\label{subsec:single}}

We begin by analyzing the action of the Kerr unitary $\hat{U}_{\rm K}(\pi/2)=e^{-i\frac{\pi}{2}\hat{n}^2}$ on a squeezed coherent state $\ket{\alpha,r}=\hat{U}_D(\alpha)\hat{U}_S(r)\ket{0}=\sum_{n=0}^\infty c_n\ket{n}$, which yields:
\begin{align}
\hat{U}_{\rm K}(\pi/2)\ket{\alpha,r}&=\sum_{n=0}^\infty c_n e^{-i\frac{\pi}{2}n^2}\ket{n}\,.
\end{align}

Separating into the even and odd subspaces, we arrive to:
\begin{equation}
    \hat{U}_{\rm K}(\pi/2)\ket{\alpha,r}=\frac{1}{\sqrt{2}}\left(e^{-i\frac{\pi}{4}}\ket{\alpha,r}+e^{i\frac{\pi}{4}}\hat{P}\ket{\alpha,r}\right)\,,
\end{equation}
where $\hat{P}=e^{i\pi\hat{n}}$ is the parity operator. Using $\hat{P}\ket{\alpha,r}=\ket{-\alpha,r}$, we obtain
\begin{equation}
\hat{U}_{\rm K}(\pi/2)\ket{\alpha,r}
=\frac{1}{\sqrt{2}}\left(e^{-i\frac{\pi}{4}}\ket{\alpha,r}
+e^{i\frac{\pi}{4}}\ket{-\alpha,r}\right)\,.
\end{equation}

Thus, the Kerr evolution splits each component into two legs with opposite displacement and a relative phase $e^{i\pi/2}$.

\subsection{Phase and amplitude propagation in the protocol~\label{subsec:propagation}}

Consider now a superposition of displaced squeezed states,
\begin{equation}
\ket{\psi}=\sum_j c_j \ket{\alpha_j,r}.
\end{equation}

Applying the Kerr unitary yields:
\begin{equation}
\hat{U}_{\rm K}(\pi/2)\ket{\psi}
=\frac{1}{\sqrt{2}}\sum_j c_j
\left(e^{-i\frac{\pi}{4}}\ket{\alpha_j,r}
+e^{i\frac{\pi}{4}}\ket{-\alpha_j,r}\right).
\end{equation}

Each leg is therefore duplicated at $\pm \alpha_j$, acquiring a relative phase and generating an interference pattern. Iterating this process across cycles leads to an exponential growth of legs with a structured phase profile. 

Let us finally note that the central leg around $\alpha=0$ is affected differently than the rest of the legs. In particular, the Kerr operation acts trivially on it, i.e., $\hat{U}_{\rm K}(\pi/2)\ket{0,r}=\ket{0,r}$. As a result, the central leg does not undergo splitting, while outer legs do. This difference induces an inhomogeneous amplitude distribution across the comb. For example, applying $\hat{U}_{\rm K}(\pi/2)$ to a two-component state yields
\begin{align}
&\hat{U}_{\rm K}(\pi/2)\left(\ket{0,r}+\ket{\alpha_0,r}\right)
=\ket{0,r}
\nonumber\\
&+\frac{1}{\sqrt{2}}\left(e^{-i\frac{\pi}{4}}\ket{\alpha_0,r}
+e^{i\frac{\pi}{4}}\ket{-\alpha_0,r}\right),
\end{align}
showing that outer legs acquire a reduced amplitude relative to the central one. These mechanisms (phase accumulation and asymmetric leg splitting) lead to the following phase and amplitude structure of the states generated by the protocol introduced in Sec.~\ref{subsec:phased-comb} at the $j$-th cycle:
\begin{equation}
\ket{\tilde{\mu}_{\mathrm{PC},j}} = \sum_{s=-s_{\rm max}}^{s_{\rm max}}
e^{i\Phi[(2s+\mu)\sqrt{\pi}]}\delta_s(\mu,s_{\rm max})\ket{(2s+\mu)\sqrt{\pi}}_{\hat{x}},
\label{eq:PCappendix}
\end{equation}
where $\Phi(x)$ is a real function determined by the circuit given by
\begin{equation}
    e^{i\Phi(x)} = \prod_{k = 1}^{n_\mathrm{cycles}}e^{i\pi\Theta{}[2^{k-(1-\mu)}-x]/2}
\end{equation}
where $\Theta(x)$ is the Heaviside function, and $\delta_s(\mu,s_{\rm max}) = \delta_{\mu,1}+\delta_{\mu,0}[(1+i)-i\delta_{s,\pm{}s_\mathrm{max}}]$, with $s_\mathrm{max}=2^{j+1}$ is the number of legs. The intuition from these formulas is to count how many times the $x$-eigenstate had been at $x<0$ during the protocol.
Note, that the amplitude difference only occurs at the boundary legs of the zero logical comb state, and thus, its importance vanishes with increasing number of cycles:
\begin{equation}
\ket{\tilde{\mu}_{\mathrm{PC},j\gg 1}}\approx  \sum_{s=-s_{\rm max}}^{s_{\rm max}} e^{i\Phi[(2s+\mu)\sqrt{\pi}]}\ket{(2s+\mu)\sqrt{\pi}}_{\hat{x}}\,,\label{eq:phasecombap}
\end{equation}

For this reason, we can say that in the large number of cycles limit, these \emph{phased-comb states} are unitarily related to standard comb states via a position-dependent phase operator $\hat{U}=e^{i\Phi(\hat{x})}$.

\section{Calculation of near-optimal channel fidelities}
\label{app:Channel_fidelity}

We use the near-optimal channel fidelity $\tilde{\mathcal{F}}_{\rm e}$ introduced in Refs.~\cite{Zheng2024,Zheng2025} as a figure of merit to characterize the error-correcting performance of the generated states. This quantity assumes an optimal recovery operation and therefore provides an upper bound on the achievable channel fidelity under a given noise model. We consider boson loss as the dominant decoherence mechanism, described by the Kraus operators
\begin{equation}
\hat{N}_k
= \sqrt{\frac{\gamma^k}{k!}}
(1-\gamma)^{\hat{n}/2} \hat{a}^k,
\end{equation}
where $k=0,\dots,\ell-1$ and $\ell$ is a truncation in the number of Kraus operators.

Given a set of logical codewords ${\{|\mu\rangle}\}$, the near-optimal channel fidelity is computed as follows:
\begin{enumerate}
\item The quantum error correction matrix is defined as
\begin{equation}
M_{\mu l, \nu k} =
\langle \mu | \hat{N}_l^\dagger \hat{N}_k | \nu \rangle.
\end{equation}

\item Since the codewords are only approximately orthogonal, we compute the overlap matrix
\begin{equation}
G_{\mu\nu} = \langle \mu | \nu \rangle,
\end{equation}
and its extension to the Kraus space,
\begin{equation}
\tilde{G}^{-1} = G^{-1} \otimes \mathbb{I}_{\ell},
\end{equation}
where $\mathbb{I}_{\ell}$ is the $\ell \times \ell$ identity matrix.

\item The near-optimal channel fidelity is then given by
\begin{equation}
\tilde{\mathcal{F}}_{\rm e} = \frac{1}{d_{\rm L}^2}
\left\|\operatorname{Tr}_{\mathrm{L}}
\left[\sqrt{\tilde{G}^{-1} M}\right]\right\|_F^2,
\end{equation}
where $d_{\rm L}=2$ and $\|\cdot\|_F$ denotes the Frobenius norm.

\end{enumerate}

\section{Phased-comb momentum eigenstates}
\label{app:momentum_eigenstates}

In this Appendix, we derive the momentum-space representation of the logical eigenstates of the infinite-energy phased-comb code. In the position basis, these states are defined as
\begin{equation}
    \begin{split}
        &\ket{+_L^\Phi} = \frac{1}{\sqrt{2}}\sum_{s\in\mathbb{Z}}e^{i\Phi(s\sqrt{\pi})}\ket{s\sqrt{\pi}}_{\hat{x}}\space{},\\
        &\ket{-_L^\Phi} = \frac{1}{\sqrt{2}}\sum_{s\in\mathbb{Z}}e^{i\Phi(s\sqrt{\pi})+is\sqrt{\pi}}\ket{s\sqrt{\pi}}_{\hat{x}}\space{}
    \end{split}
\end{equation}

We focus on $\ket{+_L^\Phi}$, as the derivation for $\ket{-_L^\Phi}$ follows analogously. Inserting the resolution of the identity in the momentum basis $\{\ket{p}\}$ and using $\langle p|x\rangle = \frac{1}{\sqrt{2\pi}}e^{-ipx}$, we obtain
\begin{equation}
\begin{split}
        \ket{+^\Phi_L} &= \frac{1}{\sqrt{2}} \int dp \ket{p} \sum_s e^{i\Phi(s\sqrt{\pi})} \langle p|s\sqrt{\pi} \rangle_{\hat{x}} \\ 
        &= \int dp \ket{p} \left[ \frac{1}{\sqrt{4\pi}} \sum_s e^{i\Phi(s\sqrt{\pi})} e^{-is\sqrt{\pi}p} \right].
\end{split}
\end{equation}

The summation can be interpreted as the Fourier transform of a phase-modulated Dirac comb. Defining
\begin{equation}
\tilde{f}(p) = \frac{1}{\sqrt{2\pi}}\int dx e^{i\Phi(x)} e^{-ipx}\,,
\end{equation}
and using that the Fourier transform of a Dirac comb yields a reciprocal comb,
\begin{equation}
\mathcal{F}\{\sum_s\delta(x-s\sqrt{\pi})\} = \sqrt{2}\sum_k\delta(p-2k\sqrt{\pi}),
\end{equation}
we obtain
\begin{equation}
 \ket{+^\Phi_L}
= \sum_{k\in\mathbb{Z}} \int \frac{dp}{\sqrt{2\pi}} \tilde{f}(p-2k\sqrt{\pi}) \ket{p}.
\end{equation}

Thus, the ideal comb structure in momentum space is replaced by a convolution with $\tilde{f}(p)$, the Fourier transform of the phase profile.

In the case of a constant phase $\Phi(x)=\phi_0$, one has $\tilde{f}(p)=\sqrt{2\pi}e^{i\phi_0}\delta(p)$, recovering the standard GKP momentum eigenstate $\ket{+_L}\propto\sum_k \ket{2k\sqrt{\pi}}_{\hat{p}}$. In contrast, for a nontrivial phase $\Phi(x)$, the function $\tilde{f}(p)$ acquires a finite width. This broadens and distorts the momentum-space leg, directly illustrating the degradation of $\hat{p}$-quadrature measurements in the phased-comb code discussed in Section~\ref{subsec:distortion}.

\section{Ancilla-based Hadamard gate~\label{app:hadamard}}

Here, we describe a phase-preserving implementation of the Hadamard gate for phased-comb states. The key challenge is that a direct Hadamard operation maps $\hat{x}\leftrightarrow\hat{p}$, which would transfer the phase dependence $e^{i\Phi(\hat{x})}$ into the momentum quadrature, leading to nonlocal phase propagation. To avoid this, we implement the Hadamard gate via a teleportation-based protocol that preserves the phase structure in the $\hat{x}$-quadrature. 

The protocol uses an ancillary qubit prepared in $\ket{+}_A$ and two entangling gates between the bosonic mode and the ancilla, namely a controlled-phase gate $\mathrm{CZ}=e^{i\hat{x}\hat{\sigma}_z}$ and a controlled-displacement (CNOT) gate $\mathrm{CNOT}=e^{i\hat{p}\hat{\sigma}_z}$, which can be implemented, e.g., using echoed conditional displacement operations~\cite{Eickbusch2022}. The circuit for the gate is illustrated below.

\begin{center}
\begin{quantikz}
\ket{+_A}&& \ctrl{1}&&\ctrl{1}&&\gate{H}&\slice{$\ket{\psi_\mathrm{out}}$} & \meter{Z} \wire[d][1]{c} \\
\ket{\psi}& &\control{}& &\targ{}&&&& \gate{U}&
\end{quantikz}
\end{center}

Considering an initial logical state in our mode, i.e., $\ket{\psi} = \alpha\ket{0_L}+\beta\ket{1_L}$, the output state after the entangling operations and the Hadamard on the ancilla reads
\begin{align}
\ket{\psi_\mathrm{out}} &=
\frac{1}{\sqrt{2}}\Big[
\ket{0_A}(\alpha\ket{+_L}-\beta\ket{-_L})\nonumber\\
&+
\ket{1_A}(\alpha\ket{-_L}+\beta\ket{+_L})
\Big].
\end{align}

Thus, depending on the measurement outcome of the ancilla, a Pauli correction is required to recover the Hadamard-transformed state:
\begin{itemize}
\item If the measurement outcome is $\ket{0}_A$, then apply $U = X$;
\item If the measurement outcome is $\ket{1}_A$, then apply $U = Z$.
\end{itemize}

This realizes the logical Hadamard transformation $\ket{0_L}\to\ket{+_L}$ and $\ket{1_L}\to\ket{-_L}$. 

Regarding the phase structure, only operations that do not commute with $\hat{x}$ can modify the phase operator $e^{i\Phi(\hat{x})}$. Among the gates in the protocol, this occurs only for the controlled-displacement gate $e^{i\hat{p}\hat{\sigma}_z}$, which induces a conditional shift of the position quadrature
\begin{equation}
e^{i\hat{p}\hat{\sigma}_z}e^{i\Phi(\hat{x})} e^{-i\hat{p}\hat{\sigma}_z}
= e^{i\Phi(\hat{x}+\hat{\sigma}_z)}\,.
\end{equation}

This shows that the phase frame acquires a shift conditioned on the ancilla state. Crucially, upon measurement of the ancilla, this shift collapses to a known classical value, and the phase dependence remains confined to the $\hat{x}$-quadrature. Therefore, the teleportation-based implementation enables a Hadamard gate that preserves the phase structure of the phased-comb encoding, avoiding the propagation of phase into the momentum quadrature.

\bibliographystyle{apsrev4-2}
\bibliography{bibliography}

\end{document}